\documentstyle[preprint,eqsecnum,aps]{revtex}
\tighten

\begin{document}
\draft

\preprint{SLAC-PUB-6218}

\title{
Magnetic Moments of the Baryon Decuplet in a Relativistic Quark Model
}
\author{Felix Schlumpf}
\address{
Stanford Linear Accelerator Center\\
Stanford University, Stanford, California 94309
}
\date{\today}
\maketitle

\begin{abstract}
The magnetic moments of the baryon decuplet are calculated in a
relativistic constituent quark model using the light-front formalism. Of
particular interest are the magnetic moments of the $\Omega^-$ and
$\Delta^{++}$ for which new recent experimental measurements are available.
Our calculation for the magnetic moment ratio $\mu(\Delta^{++})/\mu(p)$ is
in excellent agreement with the experimental ratio, while our ratio
$\mu(\Omega^-)/\mu(\Lambda^0)$ is slightly higher than the experimental
ratio.
\end{abstract}

\pacs{PACS numbers: 13.40.Fn, 12.40.Aa}

\narrowtext

\section{Introduction}\label{sec:1}
Already two of the magnetic moments of the baryon decuplet have been
measured. The most precise experiment for the $\Delta^{++}$ is a pion
bremsstrahlung analysis \cite{boss91}. The magnetic moment of the
$\Omega^-$ has been measured by the E756 collaboration \cite{dieh91}
where $\Omega^-$ hyperons are produced by a polarized neutral spin
transfer reaction. The final result from the succeeding experiment
called E800 is expected within the year \cite{rame93}. Theoretical
predictions for the $\Delta^{++}$ and $\Omega^-$ magnetic moments
have been given in many models. In particular, the simple, additive
quark model predicts the ratio of the magnetic moment of $\Delta^{++}$
and proton to be 2, and it predicts the ratio of the magnetic moment of
$\Omega^-$ and $\Lambda^0$ to be 3.

The analysis of Ref.~\cite{boss91} finds the ratio $\mu(\Delta^{++})/\mu(p)$
to be $1.62\pm 0.18$, which is significantly smaller than the additive
quark model. Most models however predict a ratio 2 or higher, a value
that is only compatible with older experiments.

We recently investigated the predictive power of a relativistic
constituent quark model formulated on the light-front \cite{schl93}. It
provides a simple model wherein we have overall an excellent and
consistent picture of the magnetic moments and the semileptonic
decays of the baryon octet. The parameters of this model
are the constituent quark mass $m$ and the scale parameter $\beta$, which
is a measure for the size of the baryon. All parameters except $\beta$
for $\Omega^-$ have been determined and fixed in Ref.~\cite{schl93}.
We predict the magnetic moments for the baryon decuplet and find
that the ratio $\mu(\Delta^{++})/\mu(p)$ is in excellent agreement with
the experiment, while our ratio $\mu(\Omega^-)/\mu(\Lambda^0)$ is
slightly higher than the experimental ratio.

In Sec.~\ref{sec:2} we give a brief summary of our model as described in
Ref.~\cite{schl93}. Section~\ref{sec:3} contains the explicit expressions
for the magnetic moments of the baryon decuplet. The numerical
results are presented in Sec.~\ref{sec:4}, and are compared with experiment
and other calculations. We summarize our investigation in a
concluding Sec.~\ref{sec:5}.

\section{Quark-model wave function}\label{sec:2}
The constituent quark model that we are going to use is described in a
previous paper \cite{schl93}. In order to fix the notation we repeat
here the essential formalism.

We shall formulate our model on the light-front which is specified by the
invariant hypersurface $x^+=x^0+x^3=0$.
The following notation is used: The
four-vector is given by $x = (x^+,x^-,x_\perp)$, where $x^\pm = x^0
\pm x^3$ and $x_\perp=(x^1,x^2)$.
Light-front vectors are denoted by boldface ${\bf x} =
(x^+,x_\perp)$, and they are covariant under kinematic Lorentz
transformations \cite{chun88}. The three momenta ${\bf p}_i$ of the quarks
can be transformed to the total and relative momenta to facilitate
the separation of the center of mass motion
\cite{bakk79}:

\begin{eqnarray}
{\bf P}&=& {\bf p}_1+ {\bf p}_2 +{\bf p}_3,
\quad \xi={p_1^+\over p_1^++p_2^+}\;,
\quad
\eta={p_1^++p_2^+\over P^+}\;,\nonumber\\
&&\\
q_\perp&=&(1-\xi)p_{1\perp}-\xi p_{2\perp}\;, \quad
Q_\perp =(1-\eta)(p_{1\perp}+p_{2\perp})-\eta p_{3\perp}\;.\nonumber
\end{eqnarray}
Note that the four-vectors are not conserved, i.e., $p_1+p_2+p_3\not= P$.
In the light-front dynamics the Hamiltonian takes the form
\begin{equation}
H={P^2_\perp + M^2 \over P^+}\;,
\end{equation}
where $M$ is the mass operator with the interaction term $W$
\begin{eqnarray}
M &=&M_0+W\;, \nonumber\\
M_0^2&=&{Q_\perp^2\over \eta(1-\eta)}+{M_3^2\over \eta}+{m_3^2\over 1-\eta},
\label{eq:2.3} \\
M_3^2&=&{q_\perp^2\over \xi (1-\xi)}+{m_1^2 \over \xi}+{m_2^2\over 1-\xi}\;,
\nonumber
\end{eqnarray}
with $m_i$ being the masses of the constituent quarks. To get a clearer
picture of $M_0$ we transform to $q_3$ and $Q_3$ by
\begin{eqnarray}
\xi&=&{E_1+q_3\over E_1+E_2}\;, \quad \eta={E_{12}+Q_3\over E_{12}+E_3}\;,
\nonumber\\
&&\\
E_{1/2}&=&({\bf q}^2+m_{1/2}^2)^{1/2}\;,\quad
E_{3}=({\bf Q}^2+m_{3}^2)^{1/2}\;,\quad
E_{12}=({\bf Q}^2+M_{3}^2)^{1/2}\;,\nonumber
\end{eqnarray}
where ${\bf q}=(q_1,q_2,q_3)$, and ${\bf Q}=(Q_1,Q_2,Q_3)$.
The expression for the mass operator is now simply
\begin{equation}
M_0=E_{12}+E_3\;, \quad M_3=E_1+E_2\;.
\end{equation}

All relevant matrix elements we investigate are
related to
\begin{equation}
\left< {\bf p}\,'\left|\bar q \gamma^+ q\right| {\bf p}\right>
\sqrt{p^{'+}p^+} \equiv {\cal M}^+ ,
\end{equation}
where the state $| {\bf p}\,\rangle\equiv|p\rangle/\sqrt{p^+}$ is
normalized according to
\begin{equation}
\left< {\bf p}\,' | {\bf p} \,\right>=\delta ({\bf p}\,' -{\bf p}\,) .
\end{equation}
The plus component of the matrix element is \cite{schl93}:
\begin{equation}
{\cal M}^+=
3{N_c\over (2\pi)^6}\int d^3qd^3Q\left({E'_3E'_{12}M\over E_3E_{12}M'}
\right)^{1/2}\Psi^\dagger({\bf q}',{\bf Q}',
\lambda')\Psi({\bf q},{\bf Q},\lambda)\;.
\label{eq:me}
\end{equation}

We can write the $\Delta^{++}$ for instance as
\begin{equation}
|\Delta^{++}\rangle = (uuu)\chi\phi .
\label{eq:wf}
\end{equation}
with $\chi$ being the spin wave function and $\phi$ being the momentum
distribution. For the latter we choose a function of $M^2_0$, in
particular we choose the same harmonic oscillator and pole type wave
function as in Ref.~\cite{schl93}:
\begin{eqnarray}
\phi_H & = & N_H \exp{(-M^2_0/2\beta^2)} ,\nonumber\\
\phi_P & = & N_P (1+M^2_0/\beta^2)^{-3.5} .
\label{eq:phi}
\end{eqnarray}
The normalization constants $N_H$ and $N_P$ are given by the constraint:
\begin{equation}
\frac{N_c}{(2\pi)^6}\int d^3q d^3Q |\phi|^2 = 1 .
\end{equation}
The spin wave function $\chi$ for the spin $\frac{3}{2}$ and
$\frac{1}{2}$ are given by
\begin{eqnarray}
\chi_\frac{3}{2} & = & \uparrow\uparrow\uparrow ,\nonumber\\
\chi_\frac{1}{2} & = & (\uparrow\uparrow\downarrow+
\uparrow\downarrow\uparrow+
\downarrow\uparrow\uparrow)/\sqrt{3} .
\end{eqnarray}
In order to get the baryon to be an eigenfunction of the spin operator we
still have to rotate the quark spins by the Melosh transformation
\cite{melo74} as follows:
\begin{equation}
\uparrow={\cal R}_i\pmatrix{1\cr 0} \hbox{  and  }
\downarrow={\cal R}_i\pmatrix{0\cr 1}\;.
\end{equation}
The Melosh rotation for three particles can be written as
\begin{eqnarray}
{\cal R}_1&=&{1\over \sqrt{a^2+Q_\perp^2}\sqrt{c^2+q_\perp^2}}
\pmatrix{ac-q_RQ_L&-aq_L-cQ_L\cr
         cQ_R+aq_R&ac-q_LQ_R}\;,\nonumber\\
{\cal R}_2&=&{1\over \sqrt{a^2+Q_\perp^2}\sqrt{d^2+q_\perp^2}}
\pmatrix{ad+q_RQ_L&aq_L-dQ_L\cr
         dQ_R-aq_R&ad+q_LQ_R}\;,\label{eq:melosh}\\
{\cal R}_3&=&{1\over \sqrt{b^2+Q_\perp^2}}\pmatrix{b&Q_L\cr
         -Q_R&b}\;,\nonumber
\end{eqnarray}
where we have defined the following quantities:
\begin{eqnarray}
a&=&M_3+\eta M\;,\quad b=m_3+(1-\eta)M\;,\nonumber\\
c&=&m_1+\xi M_3\;, \quad d=m_2+(1-\xi)M_3\;,\nonumber\\
q_R&=&q_1+iq_2\;,\quad q_L=q_1-iq_2\;,\\
Q_R&=&Q_1+iQ_2\;,\quad Q_L=Q_1-iQ_2\;.\nonumber
\end{eqnarray}

\section{Magnetic moments for the baryon decuplet}\label{sec:3}
The electromagnetic current matrix element for spin $\frac{3}{2}$
particles can be written as
\begin{equation}
\langle p',s'|J^\mu(0)|p,s\rangle=\bar u_\alpha(p',s'){\cal
O}^{\alpha\mu\beta} u_\beta(p,s) ,
\end{equation}
where $u_\alpha(p,s)$ is a Rarita-Schwinger spin-vector with momentum $p$
and spin $s$. The Lorentz covariant form for the tensor ${\cal
O}^{\alpha\mu\beta}$ may be written as
\begin{equation}
{\cal
O}^{\alpha\mu\beta}=g^{\alpha\beta}\left(F_1\gamma^\mu+\frac{F_2}{2M_B}
i\sigma^{\mu\nu}K_\nu\right)+\frac{K^\alpha K^\beta}{2M_B^2}
\left(F_3\gamma^\mu+\frac{F_4}{2M_B} i\sigma^{\mu\nu}K_\nu\right)
\end{equation}
with momentum transfer $K=p'-p$ and baryon mass $M_B$. For $K^2=0$ the
form factors $F_1$ and $F_2$ are respectively equal to the charge and the
anomalous magnetic moment, and the magnetic
moment is $\mu=F_1(0)+F_2(0)$. For this analysis we are therefore not
interested in the form factors $F_3$ and $F_4$. In order to use
Eq.~(\ref{eq:me}) we
express the form factors $F_1$ and $F_2$ in terms of the plus component
of the current:
\begin{eqnarray}
F_1(0) & = & \bigg\langle p,\frac{3}{2}\bigg|J^+\bigg|
p,\frac{3}{2}\bigg\rangle ,\nonumber\\
K_\perp F_2(0) & = & 2\left[ \sqrt{3} M_B \bigg\langle p',\frac{1}{2}
\bigg|J^+\bigg|p,\frac{3}{2}\bigg\rangle +
K_\perp \bigg\langle p,\frac{3}{2}\bigg|J^+\bigg|p,\frac{3}{2}\bigg\rangle
\right].
\label{eq:plus}
\end{eqnarray}
By inserting our wave function from Eq.~(\ref{eq:wf}) into the spin conserving
matrix element $\langle p,\frac{3}{2}|J^+|p,\frac{3}{2}\rangle$ we get
the charge of the baryon. For the spin flipping matrix element we get
\begin{equation}
\bigg\langle p',\frac{1}{2}\bigg|J^+\bigg|p,\frac{3}{2}\bigg\rangle=K_\perp f ,
\end{equation}
where $f$ is given for the different baryons as follows:
\begin{eqnarray}
f(\Delta^{++}) & = & 2 I_\Delta , \nonumber\\
f(\Delta^{+}) & = & I_\Delta , \nonumber\\
f(\Delta^{0}) & = & 0 , \nonumber\\
f(\Delta^{-}) & = & -I_\Delta , \nonumber\\
f(\Sigma^{*+}) & = & (4I_{\Sigma^*}^{(2)}-I_{\Sigma^*}^{(3)})/3 ,\nonumber\\
f(\Sigma^{*0}) & = & (I_{\Sigma^*}^{(2)}-I_{\Sigma^*}^{(3)})/3 , \nonumber\\
f(\Sigma^{*-}) & =& (-2I_{\Sigma^*}^{(2)}-I_{\Sigma^*}^{(3)})/3 ,\nonumber\\
f(\Xi^{*0}) & = & (2I_{\Xi^*}^{(3)}-2I_{\Xi^*}^{(2)})/3 , \nonumber\\
f(\Xi^{*-}) & = & (-I_{\Xi^*}^{(3)}-2I_{\Xi^*}^{(2)})/3 ,\nonumber\\
f(\Omega^-) & = & -I_\Omega .
\end{eqnarray}
The integral $I$ is given by
\begin{equation}
I=\frac{N_c}{(2\pi)^6}\int d^3q d^3Q |\phi|^2(A_1+A_2+A_3)/\sqrt{3}
\label{eq:i}
\end{equation}
where the quantities $A_i$ are
\begin{eqnarray}
A_1&=&{\eta\left( a-{Q_\perp^2\over 2(1-\eta)M}\right)\over a^2+Q_
\perp^2}{c^2\over c^2+q^2_\perp}\;,\nonumber\\
A_2&=&{\eta\left( a-{Q_\perp^2\over 2(1-\eta)M}\right)\over a^2+Q_
\perp^2}{d^2\over d^2+q^2_\perp}\;,\nonumber\\
A_3&=&{{Q_\perp^2\over 2M}-\eta b\over b^2+Q_\perp^2}\;.\nonumber
\end{eqnarray}
Note that for equal $u$ and $d$ quark masses there is an equality $A_1=A_2$
under the integral. The masses $m_i$ in our equations are set as follows
$(m=m_u=m_d)$:
\begin{eqnarray}
I_\Delta\quad :&&\quad m_1=m_2=m_3=m\;,\nonumber\\
I_{\Sigma^*}^{(2)}\quad :&&\quad m_1=m_3=m,\quad m_2=m_s\;,\nonumber\\
I_{\Sigma^*}^{(3)}\quad :&&\quad m_1=m_2=m,\quad m_3=m_s\;,\nonumber\\
I_{\Xi^*}^{(2)}\quad :&&\quad m_1=m_3=m_s,\quad m_2=m\;,\nonumber\\
I_{\Xi^*}^{(3)}\quad :&&\quad m_1=m_2=m_s,\quad m_3=m\;,\nonumber\\
I_\Omega\quad :&&\quad m_1=m_2=m_3=m_s\;.\nonumber
\end{eqnarray}
In the nonrelativistic limit, $\beta/m\to 0$, and for equal quark masses
the integral $I$ does vanish.

\section{Results and discussions}\label{sec:4}

We have calculated the magnetic moments of the decuplet baryons using
Eqs.~(\ref{eq:plus})-(\ref{eq:i}). The parameters of the model, the
constituent quark mass $m$ and the scale parameter $\beta$, have been
determined and fixed by a successful fit to the electroweak
properties of the baryon octet ($\beta_\Delta=\beta_N,
\beta_{\Sigma^*}=\beta_\Sigma, \beta_{\Xi^*}=\beta_\Xi$)
\cite{schl93}. The only new
parameter $\beta_\Omega$ is chosen to fit nicely into the raising pattern
of the $\beta$s. The parameters for both wave functions $\phi_H$ and
$\phi_P$ in Eq.~(\ref{eq:phi}) are summarized in Table~\ref{table:para}.
The corresponding results for the magnetic moments for both wave functions
(H) and (P) are given in Table~\ref{table:res} and \ref{table:compare},
together with other calculations.

In the simple nonrelativistic quark model (NQM) \cite{pdg}
the magnetic moment of a decuplet baryon is the sum of the magnetic
moments of each quark composing the baryon. This is quite different to
the fact that the magnetic moment is derived from the elastic electron
scattering at non-zero momentum transfer. The lattice result (Latt) is
taken from a recent lattice simulation of quenched QCD \cite{lein93}. The
other model calculations include results from a cloudy bag model (CB)
\cite{kriv87}, the Skyrme model (Skyr) \cite{kim89}, a Bethe-Salpeter
formalism (BS) \cite{mitr84}, an additive quark model \cite{chao90} with
effective quark masses (EM) and a calculation in which relativistic
corrections (RC) to the baryon magnetic moments are considered
\cite{geor83}. The experimental value (Expt) for the $\Delta^{++}$
is taken
from a recent pion bremsstrahlung analysis \cite{boss91}, and the
experimental value for the $\Omega^-$ is
taken from a recent investigation
where $\Omega^-$ hyperons are produced by a polarized neutral spin
transfer reaction \cite{dieh91}.

It is instructive to compare the different results of the ratios of the
magnetic moments $\mu(\Delta^{++})/\mu(p)$ and
$\mu(\Omega^-)/\mu(\Lambda^0)$. In the NQM these ratios are parameter
free and given to be 2 and 3 respectively.

The experimental value for $\mu(\Delta^{++})/\mu(p)$ of
Ref.~\cite{boss91} is lower than 2 even if we include the uncertainty of
the model dependence of the measurement $(\pm 0.16)$. Only the BS
calculation and our result are in excellent agreement with experiment. The
results from Latt and CB are even larger than 2, a result only compatible
with older experimental values \cite{nefk78,witt88}. In this sense the
magnetic moment of the $\Delta^{++}$ is a good mean to distinguish between
the different models.

The experimental value for $\mu(\Omega^{-})/\mu(\Lambda^0)$ of
Ref.~\cite{dieh91} is slightly higher than 3. It is interesting
to note that every model beyond the NQM gives also a value larger than 3,
except for EM. Our value 3.49 is closest to the central value of the
experiment, although most of the calculations are within the statistical
$(\pm 0.28)$ and systematic $(\pm 0.23)$ error of this experiment. We
could even reduce our value for the ratio $\mu(\Omega^-)/\mu(\Lambda^0)$
by using a larger value of $\beta_\Omega$ or by choosing an appropriate
anomalous magnetic moment of the strange quark.

\section{Summary}\label{sec:5}
The magnetic moments of the baryon decuplet are calculated in a
relativistic constituent quark model using the light-front
formalism. The parameters of the model are fixed by fitting the baryon
octet physics, except for $\beta_\Omega$ which is chosen in a natural
way. It is a challenge for every hadronic model to get consistent values
for the magnetic moments for both the $\Delta^{++}$ and $\Omega^-$.
Our calculation for the magnetic
moment ratio $\mu(\Delta^{++})/\mu(p)$ is in excellent agreement
with the experimental ratio, while our ratio
$\mu(\Omega^-)/\mu(\Lambda^0)$ is slightly higher than the experimental
ratio.

\acknowledgments

It is a pleasure to thank W.~Jaus for helpful discussions. I am grateful
to X.~Ji for his hospitality at M.I.T. where the draft of this paper was
written.
This work was supported in part by the Schweizerischer Nationalfonds and
in part by the Department of Energy, contract DE-AC03-76SF00515.

\narrowtext
\begin{table}
\caption{The parameter of the constituent quark model for the harmonic
oscillator wave function (H) and for the pole type (P) wave function.
Both the quark masses $m$ and the scale parameter $\beta$ are given in
units of GeV.}
\begin{tabular}{cdd}
Parameters & H & P \\
\tableline
$m_u=m_d$ & 0.26 & 0.263 \\
$m_s$ & 0.38 & 0.38 \\
$\beta_\Delta$ & 0.55 & 0.607 \\
$\beta_{\Sigma^*}$ & 0.60 & 0.75 \\
$\beta_{\Xi^*}$ & 0.62 & 0.90 \\
$\beta_\Omega$ & 0.70 & 1.05 \\
\end{tabular}
\label{table:para}
\end{table}

\narrowtext
\begin{table}
\caption{Magnetic moments of the baryon decuplet. The calculations of the
present work with the harmonic oscillator wave function (H) and the pole
type wave function (P) are compared with the simple nonrelativistic quark
model (NQM), with a lattice calculation (Latt) and with the Skyrme model
(Skyr). The magnetic moments are given in units of the nuclear magneton.
References are given in the text.}
\begin{tabular}{lddddd}
Baryon & H & P & NQM & Latt & Skyr \\
\tableline
$\Delta^{++}$ & 4.76 & 4.93 & 5.56 & 4.91 & 4.53 \\
$\Delta^{+}$ & 2.38 & 2.47 & 2.73 & 2.46 & 2.09 \\
$\Delta^{0}$ & 0.00 & 0.00 & --0.09 & 0.00 & --0.36 \\
$\Delta^{-}$ & --2.38 & --2.47 & --2.92 & --2.46 & --2.80 \\
$\Sigma^{*+}$ & 1.82 & 1.84 & 3.09 & 2.55 & 2.55 \\
$\Sigma^{*0}$ & --0.27 & --0.28 & 0.27 & 0.27 & --0.02 \\
$\Sigma^{*-}$ & --2.36 & --2.41 & --2.56 & --2.02 & --2.60 \\
$\Xi^{*0}$ & --0.60 & --0.56 & 0.63 & 0.46 & 0.40 \\
$\Xi^{*-}$ & --2.41 & --2.41 & --2.20 & --1.68 & --2.31\\
$\Omega^-$ & --2.48 & --2.47 & --1.84 & --1.40 & --1.98 \\
\end{tabular}
\label{table:res}
\end{table}

\mediumtext
\begin{table}
\caption{Comparison of our calculations (H) and (P)
of the magnetic moments for the
$\Delta^{++}$ and $\Omega^-$ with other calculations and experiment
(Expt). The calculations are the simple nonrelativistic quark model (NQM),
lattice calculations (Latt), Skyrme model (Skyr), cloudy bag model (CB),
Bethe-Salpeter formalism (BS), an additive quark model based on effective
quark masses (EM) and a calculation including relativistic corrections
(RC). The experimental value given for $\Delta^{++}$ has some model
dependence. All numbers are given in units of the nuclear magneton.
References are given in the text.}
\begin{tabular}{ld@{${}\pm{}$}ldddddddddd}
Magnetic Moment & \multicolumn{2}{c}{Expt} & H & P & NQM & Latt &
 Skyr & CB & BS & EM & RC \\
\tableline
$\mu(\Delta^{++})$ & 4.52 & 0.50 & 4.76 & 4.93 & 5.56 & 4.91 & 4.53 &
6.54 & 4.44 & -- & -- \\
$\mu(\Delta^{++})/\mu(p)$ & 1.62 & 0.18 & 1.69 & 1.75 & 2.00 & 2.18 &
1.98 & 2.34 & 1.59 & -- & -- \\
$\mu(\Omega^{-})$ & --1.94 & 0.17 & --2.41 & --2.41 & --1.84 & --1.40 &
--1.98 & --2.52 & -- & --1.69 & --2.25 \\
$\mu(\Omega^{-})/\mu(\Lambda^0)$ & 3.16 & 0.28 & 3.49 & 3.49 & 3.00 & 3.6 &
3.73 & 4.13 & -- & 2.77 & 3.66 \\
\end{tabular}
\label{table:compare}
\end{table}


\begin{thebibliography}{10}

\bibitem{boss91}
A.~Bosshard {\it et~al.}, Phys. Rev. D {\bf 44}, 1962 (1991).

\bibitem{dieh91}
H.~T. Diehl {\it et~al.}, Phys. Rev. Lett. {\bf 67}, 804 (1991).

\bibitem{rame93}
G.~Rameika, Fermilab Report Jan. Feb. March, 5 (1993).

\bibitem{schl93}
F. Schlumpf, Phys. Rev. D {\bf 47},  4114  (1993).

\bibitem{chun88}
P.~L. Chung {\it et~al.}, Phys. Rev. C {\bf 37},  2000  (1988).

\bibitem{bakk79}
B.~L.~G. Bakker, L.~A. Kondratyunk, and M.~V. Terent'ev, Nucl. Phys. B {\bf
  158}, 497  (1979).

\bibitem{melo74}
H.~J. Melosh, Phys. Rev. D {\bf 9},  1095  (1974).

\bibitem{pdg}
Particle Data Group, Phys. Rev. D {\bf 45},  1  (1992).

\bibitem{lein93}
D.~B. Leinweber, T.~Draper, and R.~M. Woloshyn, University of Maryland,
Report No. MDDP-PP-92-188 (1992).

\bibitem{kriv87}
M.~I. Krivoruchenko, Sov. J. Nucl. Phys. {\bf 45}, 109 (1987).

\bibitem{kim89}
J.~H. Kim, C.~H. Lee, and H~.K. Lee, Nucl. Phys. {\bf A501}, 835 (1989).

\bibitem{mitr84}
A.~Mitra and A.~Mittal, Phys. Rev. D {\bf 29}, 1399 (1984).

\bibitem{chao90}
K.~T. Chao, Phys. Rev. D {\bf 41}, 920 (1990).

\bibitem{geor83}
H.~Georgi and A.~Manohar, Phys. Lett. {\bf 132B}, 183 (1983).

\bibitem{nefk78}
B.~M.~K. Nefkens {\it et~al.}, Phys. Rev. D {\bf 18}, 391 (1978).

\bibitem{witt88}
R.~Wittman, Phys. Rev. D {\bf 37}, 2075 (1988).

\end{thebibliography}
\end{document}